\documentclass[aps,prc,twocolumn,superscriptaddress]{revtex4-1}
\usepackage{graphicx}
\usepackage{amsmath}
\usepackage{braket}
\usepackage{url}
\usepackage{ulem}
\usepackage{multirow}
\usepackage{amssymb} 
\usepackage{booktabs}
\usepackage{rotating}
\usepackage{bm}
\usepackage{bbm}
\usepackage{longtable}
\usepackage{subfigure}
\usepackage{tikz}
\usetikzlibrary{matrix}

\usepackage[colorlinks,
linkcolor=blue,
anchorcolor=red,
urlcolor=red,
citecolor=red]{hyperref}

\begin{document}

\title{Spectroscopic factor calculations in the \textit{ab initio} no-core shell model}

\author{M. R. Xie}
\author{J. G. Li}\email[]{jianguo\_li@impcas.ac.cn}
\author{N. Michel}
\author{W. Zuo}
\affiliation{Institute of Modern Physics, Chinese Academy of Sciences, Lanzhou 730000, China}
\affiliation{School of Nuclear Science and Technology, University of Chinese Academy of Sciences, Beijing 100049, China}

\date{\today}

\begin{abstract}
The convergence properties of spectroscopic factors in the \textit{ab initio} no-core shell model are hereby investigated.
For this, we consider nuclear energies and spectroscopic factors in $A = 6$ and 7 isotopes, using the chiral forces NNLO$_{\rm opt}$ and N$^3$LO. While low-lying spectrum energy demonstrates remarkable convergence with the increase of model space within the no-core shell model, the spectroscopic factor exhibits no definitive convergence trend and seems independent of the employed nuclear interaction.
The use of spectroscopic factors issued from the no-core shell model to calculate cross-sections of knockout reaction might then be questionable.
The results are compared with that of the standard shell model and \textit{ab initio} Monte Carlo calculations.


\end{abstract}


\maketitle

\section{Introduction}

In nuclear physics, spectroscopic factors (SFs) serve as vital keys to unlocking insights into the nuclear structure and behavior~\cite{AUMANN2021103847, PhysRevLett.102.062501, PhysRevLett.130.172501, PhysRevLett.129.152501, PhysRevLett.126.082501, BECK2007128}. These factors quantify the likelihood of a nucleon occupying a particular single-particle state within the nucleus~\cite{PhysRevLett.126.082501, BECK2007128,CHEN2018412}, influencing an array of nuclear reactions~~\cite{AUMANN2021103847}. 
SFs pertaining to transfer and knockout nuclei have long been at the forefront of experimental investigations in nuclear physics~\cite{AUMANN2021103847, PhysRevLett.102.062501, PhysRevLett.130.172501, PhysRevLett.129.152501, PhysRevLett.126.082501, BECK2007128, PhysRevC.78.041302, PhysRevC.104.014310, PhysRevC.104.064605}.
Moreover, they are of paramount importance in nuclear astrophysics~\cite{10.3389/fphy.2020.602920}, where they aid in comprehending stellar nucleosynthesis and related processes. 
When applied to nucleon transfer or knockout reactions, modern nuclear reaction theories often produce results that significantly deviate from experimental data~\cite{AUMANN2021103847, PhysRevC.90.057602, PhysRevC.103.054610, PhysRevC.105.024613}. 
In experiments, SF is often deduced from direct reaction measurements, where it acts as a normalization factor in the reaction cross-section~\cite{AUMANN2021103847}. However, the SF is always calculated using the nuclear structure models, and the calculated value also depends on the specific model adopted~\cite{AUMANN2021103847, PhysRevC.103.054610, PhysRevC.105.024613, XIE2023137800}.
It has been observed that the SF extracted from the ($e,e'p$) reaction is typically $30\%-40\%$ smaller compared to predictions based on the mean-field shell model~\cite{LAPIKAS1993297, KRAMER2001267}. This discrepancy is often referred to as ``quenching''. The origin of this ``quenching'' is currently a topic of intense research~\cite{AUMANN2021103847}. It is speculated that it may stem from the complexities of long-range and short-range correlations~\cite{AUMANN2021103847, PhysRevLett.130.172501}. However, the exact mechanisms and implications of these correlations remain an open question, prompting further investigations.
In nuclear structure, the precision in determining SF is pivotal for elucidating a variety of phenomena~\cite{PhysRevLett.103.202502, PhysRevLett.107.032501, PhysRevC.103.054610, PhysRevC.105.024613, XIE2023137800}.

In recent years, \textit{ab initio} calculations have emerged as pivotal tools in advancing the field of nuclear physics~\cite{BARRETT2013131, RevModPhys.87.1067, HERGERT2016165, Hagen_2016, PhysRevC.100.054313,doi:10.1146/annurev-nucl-101917-021120,PhysRevC.107.014302}.
One of the main advantages of these \textit{ab initio} calculations is their inherent objectivity, stemming from their independence from empirical parameters and experience rules.
Furthermore, \textit{ab initio} calculations offer a refined avenue for treating internucleon correlations precisely, providing predictive insights into the properties of nuclei that are either challenging to probe or currently beyond the reach of experimental investigations~\cite{PhysRevLett.130.172501, PhysRevC.105.024613, PhysRevC.100.054313}.
Among various methods, the \textit{ab initio} no-core shell model (NCSM) is particularly notable~\cite{BARRETT2013131, PhysRevC.98.014002, PhysRevC.77.024301, PhysRevC.79.064324, PhysRevC.69.014311, PhysRevC.104.024319}.
A distinguishing feature of the NCSM is its equal treatment of every nucleon within the nucleus without the presumption of an inert core. This distinction means the NCSM provides a more comprehensive and democratic picture of nuclear behavior.
The main challenge facing the \textit{ab initio} calculations in this field is the considerable computational cost associated with these calculations~\cite{PhysRevC.104.054315, PhysRevC.79.014308, PhysRevC.99.054308}.
Strategies, including renormalization and extrapolation methods, have been increasingly adopted to address this challenge with notable success~\cite{BOGNER200821, BOGNER201094, PhysRevC.92.024003, PhysRevC.77.024301, PhysRevC.79.014308, PhysRevC.79.064324,PhysRevC.102.024305,KNOLL2023137781,PhysRevC.105.L061303}. {The energies, electromagnetic, and radius have been systematically investigated by the \textit{ab initio} NCSM calculation. However, the systematic investigation of the convergence of SF with \textit{ab initio} NCSM calculation is lacking.}

In this work, we employ the \textit{ab initio} NCSM to calculate the SFs. We begin by briefly introducing the NCSM framework and the definitions of the overlap function and the SF. Subsequently, we utilize the NCSM to calculate the energies of {$^6$He, $^6$Li, and $^7$Li, and discussed the convergence with the increasing model space.
We then systematically calculate the SFs of $^7$Li and present their convergence with the increasing model space.}
In our practical application, we use the calculated SFs to compute the cross-section of knockout reactions, and the results are compared with the calculations with the SFs derived from the standard shell model (SM) and \textit{ab initio} Monte Carlo methods.
Finally, a summary will be provided.

\section{Method}
\subsection{No core-shell model}
The nucleus is an intricate system composed of $A$-nucleons, consisting of neutrons and protons. The corresponding Hamiltonian of this $A$-body system can be expressed as
\begin{eqnarray}
\label{NCSM}
\hat{H} = \sum_{i < j} \frac{\left(p_{i}-p_{j}\right)^{2}}{2 A m}+\widehat{V}_{NN}+..., 
\end{eqnarray}
where $m$ denotes the nucleon mass and $\widehat{V}_{NN}$ represents the nucleon-nucleon ($NN$) interaction, to which three-, four-, ... body forces should be considered in principle. However, only the $NN$ is considered in the present work. 
Moreover, the Coulomb force is also included in the practical calculations. Two sets of chiral interactions, the optimized next-to-next-to-leading order (NNLO$_{\rm opt}$)~\cite{PhysRevLett.110.192502} and the next-to-next-to-next-to-leading order (N$^3$LO) interaction~\cite{PhysRevC.68.041001},  are adopted.
To accelerate the convergence of the many-body calculations, the chiral N$^3$LO potential is softened by the similarity renormalization group (SRG) method with $\lambda$ = 2.0 fm$^{-1}$~\cite{PhysRevC.75.061001}. The bare forces are used in practical calculations for the NNLO$_{\rm opt}$ interaction.

Within the NCSM framework, the many-body Schr\"{o}dinger equation is solved in the harmonic-oscillator (HO) basis, which is the linear combination of the Slater determinant of the single-particle states. In real calculations, a large but finite HO basis is utilized.
The results of the NCSM calculations rely on two pivotal \textit{parameters}: the frequency of the HO basis ($\hbar w$) and the truncation of the model space ($N_{\rm max}$)~\cite{BARRETT2013131}. The $N_{\rm max}$ presents the total number of oscillator quanta allowed above the minimum for a given nucleus in the many-body HO basis space. Our objective is to achieve convergence in this two-dimensional \textit{parameter} space ($\hbar w$, $N_{\rm max}$), where convergence implies the results are independent of both $\hbar w$ and $N_{\rm max}$, within estimated uncertainties. 
The dimensionality of the NCSM calculations increases drastically with the mass of nuclei.
Due to the limitations of current large supercomputers, the NCSM can only give near-convergence results for nuclei with nucleon number $A \le 16$~\cite{BARRETT2013131, PhysRevC.106.064002}. 
Balancing the quest for accurate results with the constraints of computational resources, the NCSM calculations are calculated within the model space with $N_{\rm max}$, ranging from 0 to 12 for the $A=6$ and 7 isotopes. Simultaneously, the  $\hbar w$ varies from 10 MeV to 32 MeV.

\subsection{Overlap function and spectroscopic factor}

The overlap function, denoted as $I_{\ell j}(r)$, is defined as follows
\begin{eqnarray}
\label{overlap}
I_{\ell j}(r) & = & \langle \Psi_A^{J_A}| \left [|\Psi_{A-1}^{J_{A-1}}\rangle \otimes |r \ell j\rangle^{J_A} \right]\rangle, \nonumber \\
          & = & \frac{1}{\sqrt{2J_A + 1}} \sum_i \langle \Psi_A^{J_A}||a_{n_i \ell j}^{\dagger}|| \Psi_{A-1}^{J_{A-1}}\rangle  u_{n_i}(r),
\end{eqnarray}
where $|\Psi^{J_A}_A\rangle$ and $|\Psi^{J_{A-1}}_{A-1}\rangle$ are the wave functions of the $A$ and $A-1$ nuclear systems, respectively.
The variables $\ell$ and $j$ denote the orbital and total angular momenta of the partial wave being considered in $I_{\ell j}(r)$. 
 And $a_{n_i \ell j}^{\dagger}$ is the creation operator associated with the $|n_i \ell j \rangle$ state; 
$u_{n_i}(r)$ represents the radial wave function of the $|n_i \ell j \rangle$ state.

The SF is defined as the norm of the radial overlap function $I_{\ell j}(r)$ from Eq.~(\ref{overlap}). This can be represented as
\begin{equation}
\label{specfac}
{C^2 S_{\ell j} = \int I_{\ell j}(r)^2 dr},
\end{equation} 
where $C^2 S$ is a standard notation for SF.
Current nuclear reaction theory defines the nuclear knockout reaction cross-section as
\begin{eqnarray}
\label{eq-sigma}
\sigma_{\rm total}=\sum_{ \ell j} f_\text{CoM} C^{2} S( \ell j) \sigma_{sp}\left( \ell j\right),
\end{eqnarray}
where the total cross-section is represented by $\sigma_{\rm total}$.
The term  $\sigma_{sp}\left( \ell j\right)$ denotes the cross-section of the removal of a nucleon in the $\ell j$ partial wave.
The $f_\text{CoM}$ is the center-of-mass (CoM) correction factor, which is equal to 1 in NCSM and  $(A/A-1)^N$ in standard SM calculations, respectively, where $N$ is the major oscillator quantum number.
$\sigma_{sp}\left( \ell j\right)$ is calculated using reaction models, such as the 
eikonal model~\cite{doi:10.1146/annurev.nucl.53.041002.110406}.
By combining the calculated $\sigma_{s p}\left( \ell j\right)$ with the $C^2S$ obtained from nuclear many-body models for structure, it becomes feasible to determine the theoretical cross-section of the knockout reaction. This procedure is paramount as it establishes a crucial bridge linking theoretical predictions directly with experimental data. 

Let us comment on the CoM effects in Eqs.~(\ref{overlap},\ref{specfac},\ref{eq-sigma}) in NCSM. Even though $| \Psi_{A-1}^{J_{A-1}} \rangle$ and $| \Psi_A^{J_A} \rangle$ are products of CoM times intrinsic wave functions, i.e.~they are free from CoM spuriosity, their CoM parts are different, on the one hand, and the operators in Eqs.~(\ref{overlap},\ref{specfac}) are not translationally invariant, on the other hand. Consequently, CoM spuriosity is, in principle, present in Eqs.~(\ref{overlap},\ref{specfac}).
However, SF and overlap functions are not physical observables, so an exact removal of CoM effects is not meaningful even in NCSM. The only physical quantity arising from overlap functions and entering cross-section expressions is the asymptotic normalization coefficient \cite{ANC_review, PhysRevC.56.1302}. The latter is defined in the asymptotic region, where CoM corrections vanish. In fact, Eq.~(\ref{eq-sigma}) is an approximate expression, whereby the use of asymptotic normalization coefficients or SFs should be equivalent (up to a different normalization of $\sigma_{sp}\left( \ell j\right)$ \cite{ANC_review, PhysRevC.56.1302}), so that it is legitimate to ignore CoM corrections in Eqs.~(\ref{overlap},\ref{specfac},\ref{eq-sigma}) with NCSM.


\section{Results} 

\begin{figure}
    \centering    
    \includegraphics[width=0.4\paperwidth]{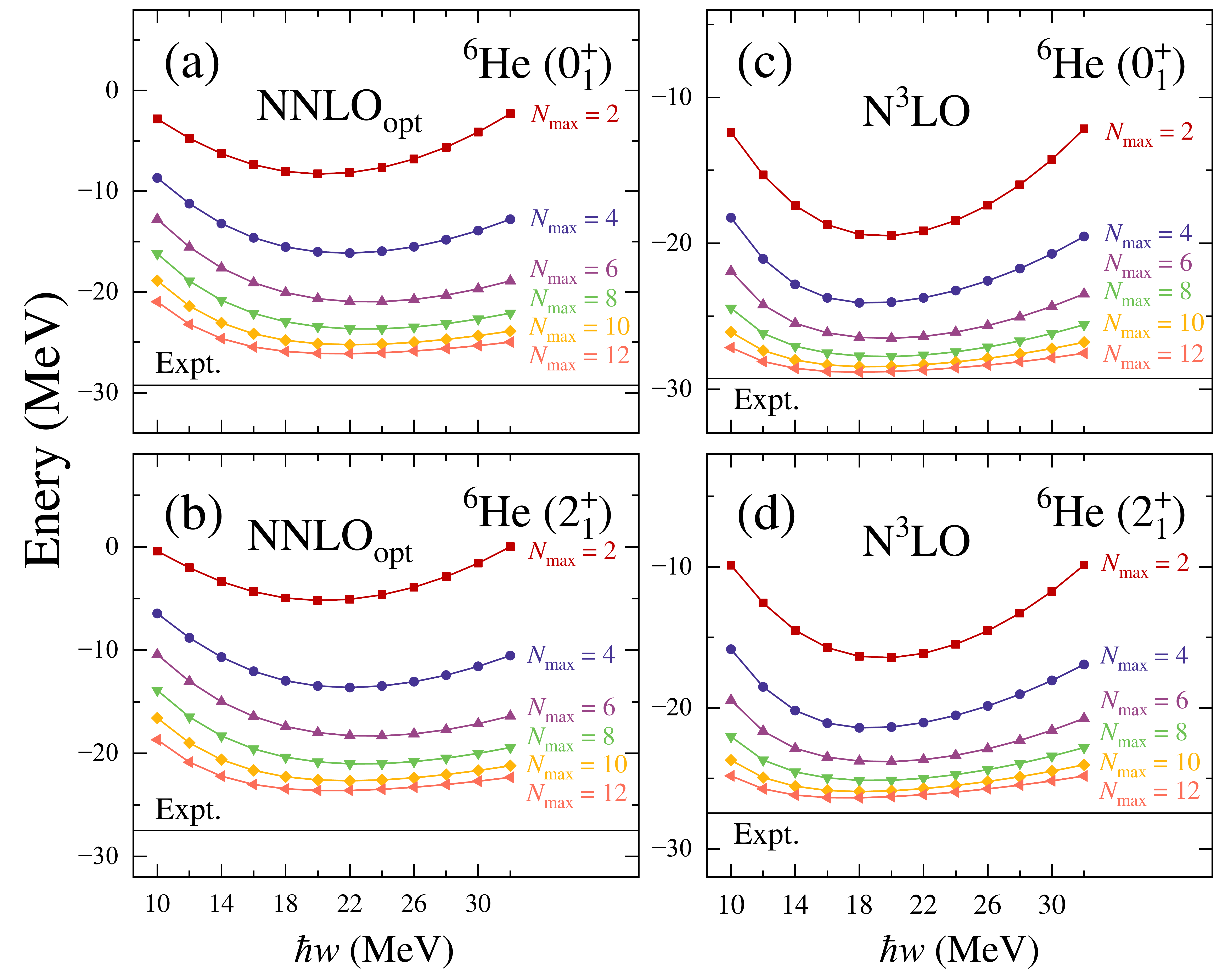}
    \caption{The calculated energy of the ground state (0$_1^+$) and excited state (2$_1^+$) for $^6$He with NNLO$_{\rm opt}$ and N$^3$LO interaction, plotted as functions of truncation of model space $N_{\rm max}$ and HO frequency $\hbar w$. The chiral N$^3$LO interaction is renormalized via the SRG method with $\lambda$ = 2.0 fm$^{-1}$. Black lines correspond to experimental data. 
    }
    \label{6He}
\end{figure}

\begin{figure}
    \centering    
    \includegraphics[width=0.4\paperwidth]{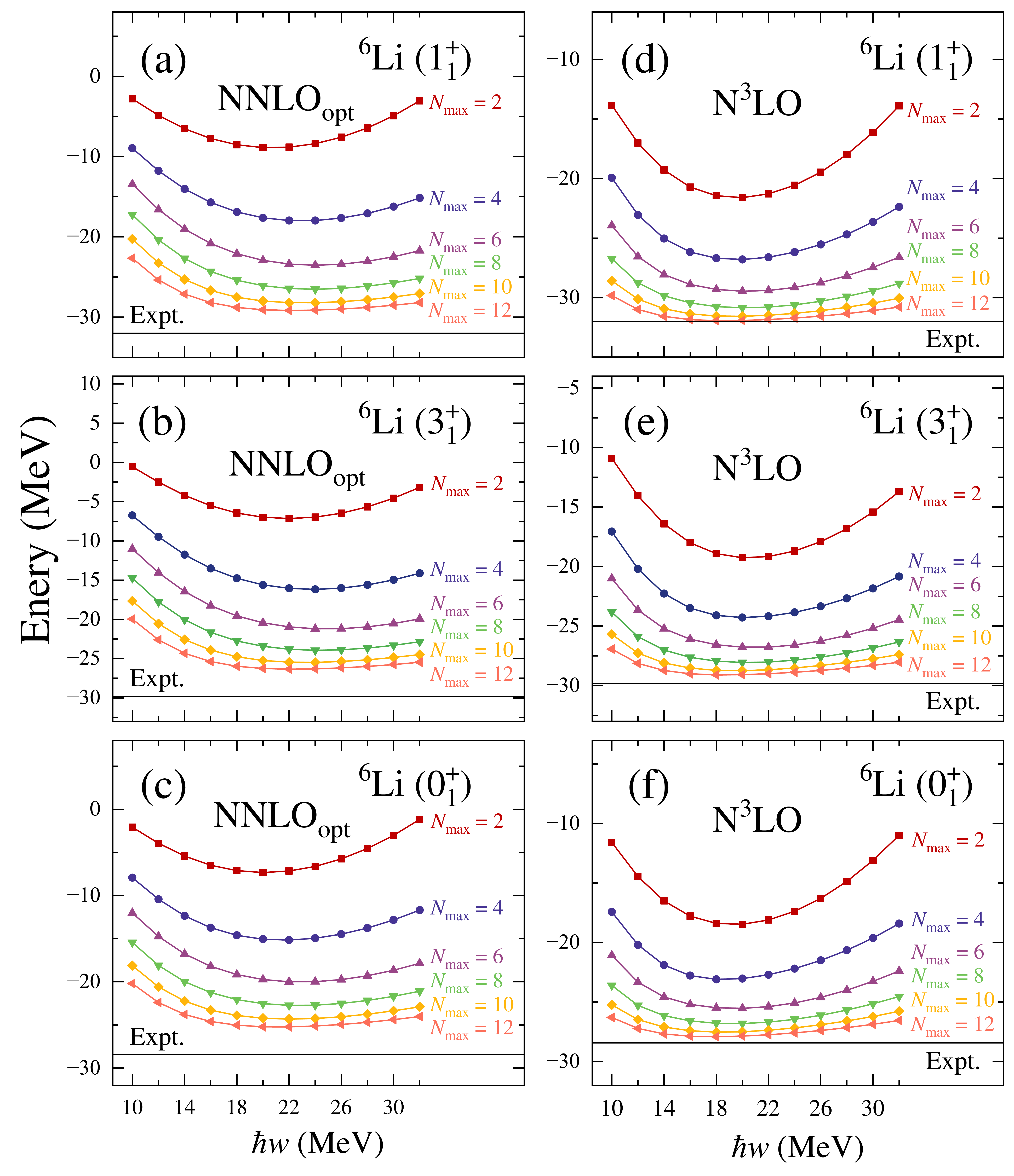}
    \caption{Similar to Fig.~\ref{6He}, but for the ground state (1$_1^+$) and excited state (3$_1^+$, 0$_1^+$) of $^6$Li. 
    }
    \label{6Li}
\end{figure}

\begin{figure}
    \centering
\includegraphics[width=0.4\paperwidth]{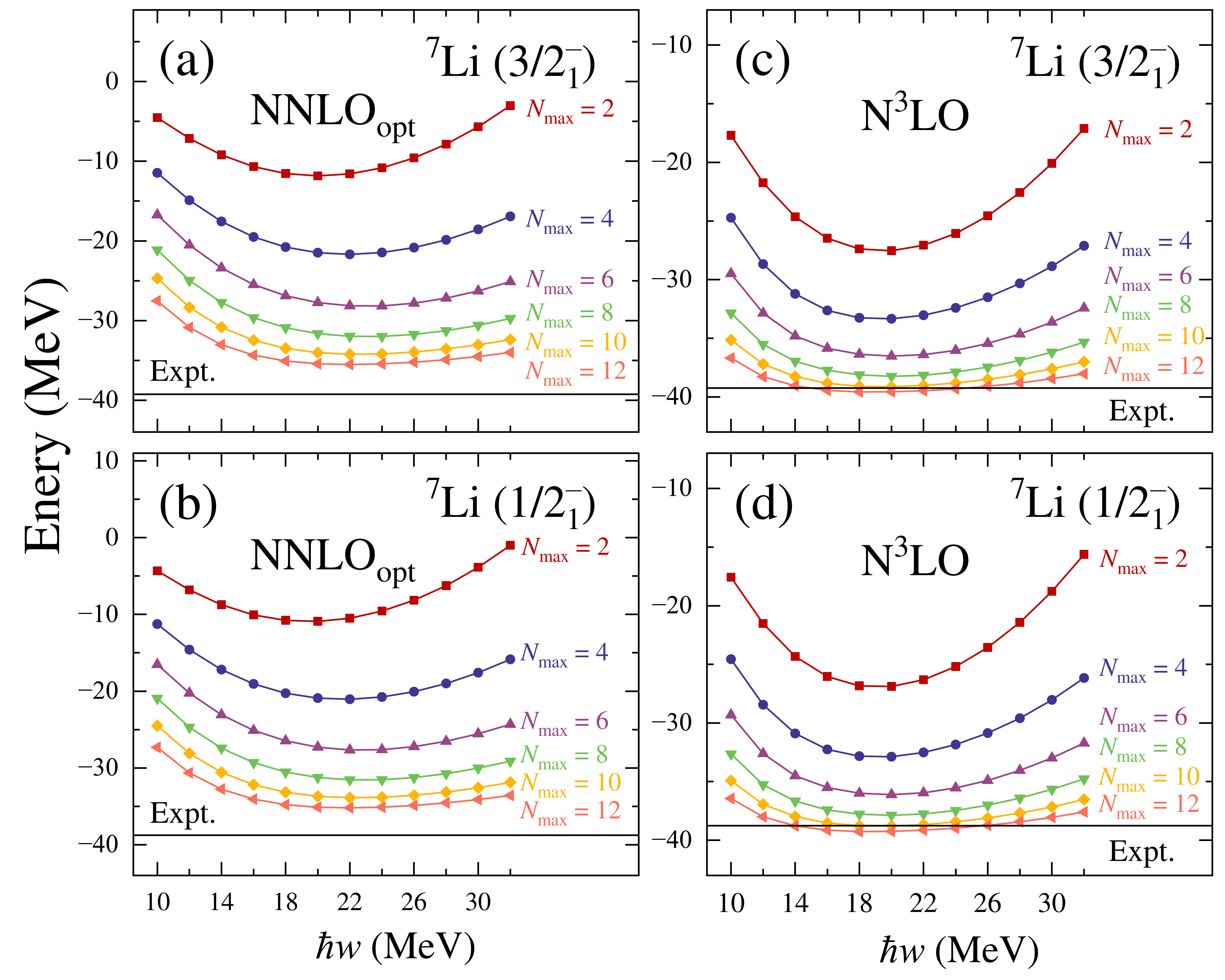}
    \caption{Similar to Fig.~\ref{6He}, but for the ground state (3/2$_1^-$) and excited state (1/2$_1^-$) of $^7$Li.}\label{7Li}
\end{figure}

{Firstly, we focus on the energy of $^{6}$He, $^{6}$Li, and $^7$Li nuclei.} The energies of the ground state and excited state in those nuclei are calculated using NCSM with NNLO$_{\rm opt}$ and N$^3$LO interactions as a function of $N_{\rm max}$ and $\hbar w$. {Results are presented in Figs.~\ref{6He},\ref{6Li},\ref{7Li}.}
A point of specific interest here is the dependency of convergence properties in model space truncation $N_{\rm max}$ and HO frequency $\hbar w$ in NCSM calculations. 
As the figures illustrate, energy dependency on $\hbar w$ reduces as $N_{\rm max}$ increases.
Ideally, the computed observable should be independent of the parameters~\cite{PhysRevC.86.031301}. Moreover, the convergence rate with $N_{\rm max}$ is different for different states~\cite{PhysRevC.69.014311} and the minimum energy value is observed to lie between $\hbar w = 18 - 22$ MeV.
Another noteworthy observation is that as $N_{\rm max}$ increases, the gap between different curves narrows, indirectly indicating a convergence trend in the energy calculation results.


We observe that the results of N$^3$LO and NNLO$_{\rm opt}$ have distinct differences in energy spectra and convergence rates. 
The calculations employing N$^3$LO interaction demonstrate lower energy and notably faster convergence than those using NNLO$_{\rm opt}$ interaction.
Moreover, the proximity of the results for different $\hbar w$ values indicates that N$^3$LO exhibits lesser parameter dependence with a fixed largest $N_{\rm max}$ in the practical calculation.
The energy of infinite model spaces ($N_{\rm max} \to \infty$) can be extrapolated based on the results of calculations in finite model space using extrapolation formulas~\cite{PhysRevC.86.054002, PhysRevC.91.061301}. 
In this work, the extrapolation is done based on a simple exponential form, written as~\cite{PhysRevC.104.054315, PhysRevC.105.L061303}
\begin{equation}
\label{eq9}
E(N_{\rm max}) = E(N_{\rm max} \to \infty) + A_{1} \exp(-A_{2} \cdot N_{\rm max}),
\end{equation}
where $E(N_{\rm max} \to \infty)$, $A_{1}$, and $A_2$ represent specific parameters that can be obtained by fitting the NCSM calculations in finite model space.

As depicted in Fig.~\ref{6He}, we present the calculated results for $^6$He. 
The lowest energy obtained in the NCSM calculations of $^{6}$He in the $N_{\rm max} =12$ model space is at $\hbar w = 18$ MeV for N$^3$LO interaction and at $\hbar w = 22$ MeV for NNLO$_{\rm opt}$ interaction. 
{For the ground state energy of $^{6}$He, the extrapolation results are $-27.378$(042) and $-29.246$(040) MeV for NNLO$_{\rm opt}$ and N$^3$LO,} respectively, and the experimental data of ground state energy of $^{6}$He is $-29.271$(054) MeV~\cite{ensdf}. 
The error of the energy is deduced from the uncertainty of extrapolated energies based on the NCSM calculations using $N_{\rm max}=2$, 4, 6, 8, 10, and 12 model spaces.
For the $2_1^+$ excited state, a comparable situation occurs where the result of N$^{3}$LO is closer to experimental data with respect to that of  NNLO$_{\rm opt}$. 
Similar results have also been obtained for the low-lying states in $^{6}$Li and $^{7}$Li, see Table~\ref{energy} for detail.
\begin{table}[]
\caption{Energies (in MeV) of the low-lying states in $^6$He and $^{6,7}$Li  extrapolated from NCSM calculations. The $\rm NNLO_{opt}$ and $\rm N^3LO$ interactions were used with $\hbar w=22$ and 18 MeV, respectively. Experimental data are taken from Ref.~\cite{ensdf}.}
\begin{tabular}{cccccc}
\hline\hline
 Nucleus & State & $\rm NCSM_{NNLO_{opt}}$ & $\rm NCSM_{N^3LO}$ & Expt.~\cite{ensdf}  \\
 \hline\hline
 $^6$He & $0_1^+$ & $-27.378(042)$ & $-29.246(040)$ & $-29.271(054)$ \\
 $^6$He & $2_1^+$ & $-24.757(112)$ & $-26.814(100)$ & $-27.474(025)$ \\
 $^6$Li & $1_1^+$ & $-30.593(072)$ & $-32.392(038)$ & $-31.994(018)$ \\
 $^6$Li & $3_1^+$ & $-27.527(058)$ & $-29.518(065)$ & $-29.808(002)$  \\
 $^6$Li & $0_1^+$ & $-26.574(076)$ & $-28.348(041)$ & $-28.431(010)$  \\
 $^7$Li & $3/2_1^-$ & $-37.777(196)$ & $-40.195(034)$ & $-39.245(042)$  \\
 $^7$Li & $1/2_1^-$ & $-37.419(137)$ & $-39.894(033)$ & $-38.767(003)$  \\
 \hline\hline
 \label{energy}
\end{tabular}
\end{table}

\begin{figure*}
    \centering
\includegraphics[width=0.75\paperwidth]{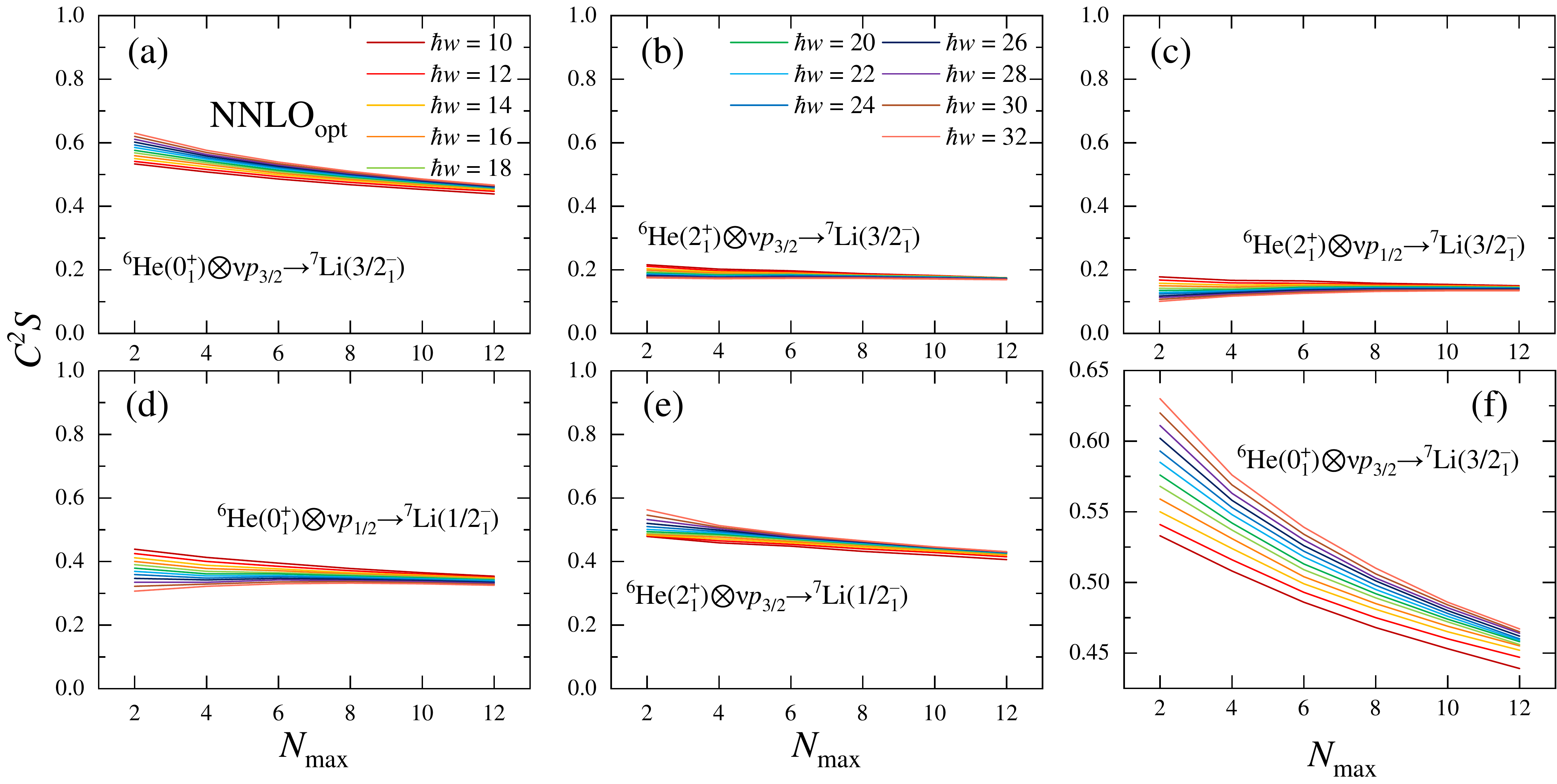}
    \caption{The calculated SFs of $^6$He $\otimes$ $p_{3/2,1/2}$ $\to$ $^7$Li using NCSM with NNLO$_{\rm opt}$ interaction as functions of $N_{\rm max}$ and $\hbar w$.}
    \label{6He7Li_NNLO}
\end{figure*}

\begin{figure*}
    \centering
\includegraphics[width=0.75\paperwidth]{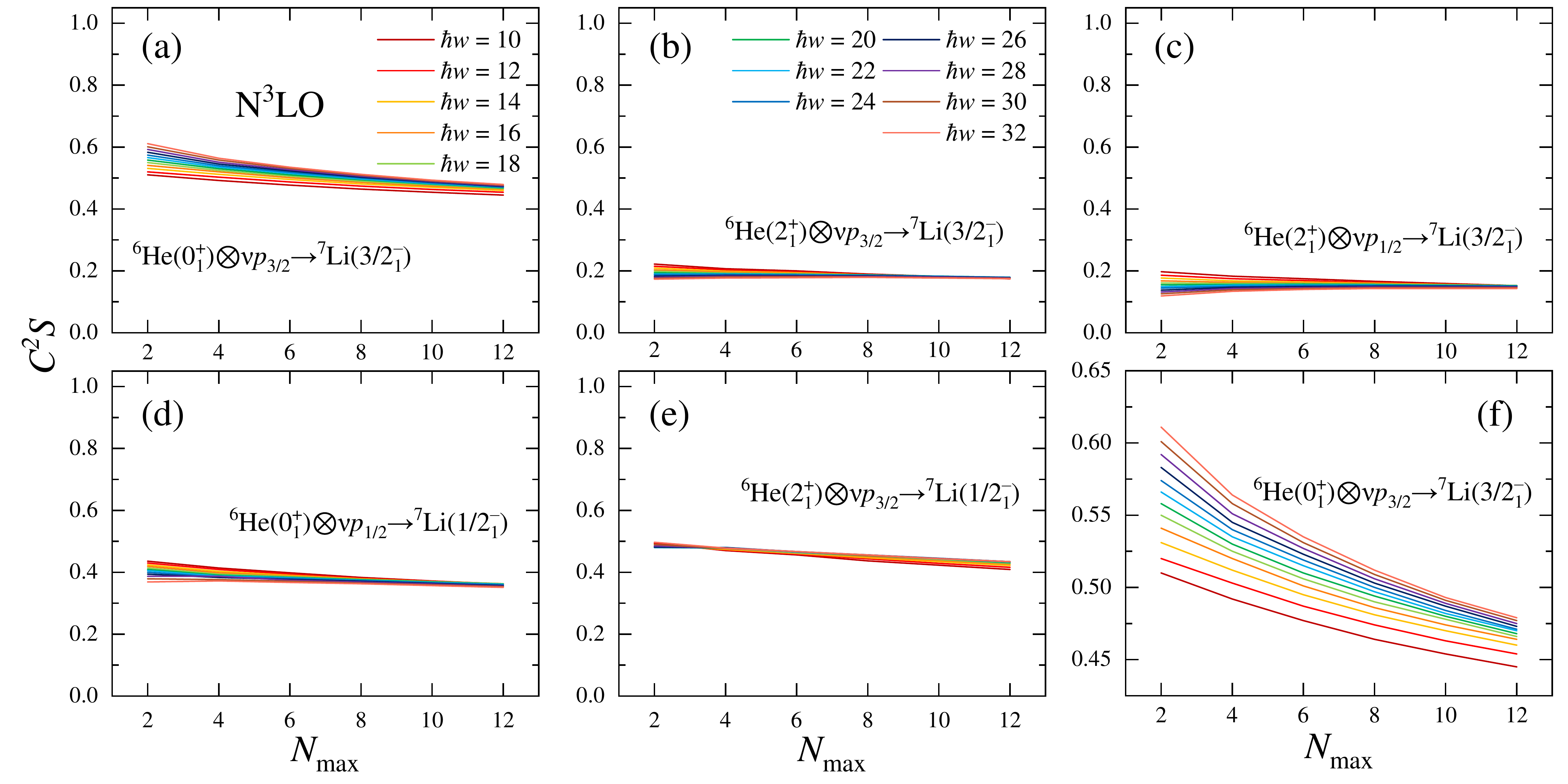}
    \caption{Similar to Fig.~\ref{6He7Li_NNLO}, but for $^6$He $\otimes$ $p_{3/2,1/2}$ $\to$ $^7$Li using chiral N$^3$LO interaction renormalized via  SRG with $\lambda$ = 2.0 fm$^{-1}$.}
    \label{6He7Li_SRG}
\end{figure*}

\begin{figure*}
    \centering
\includegraphics[width=0.75\paperwidth]{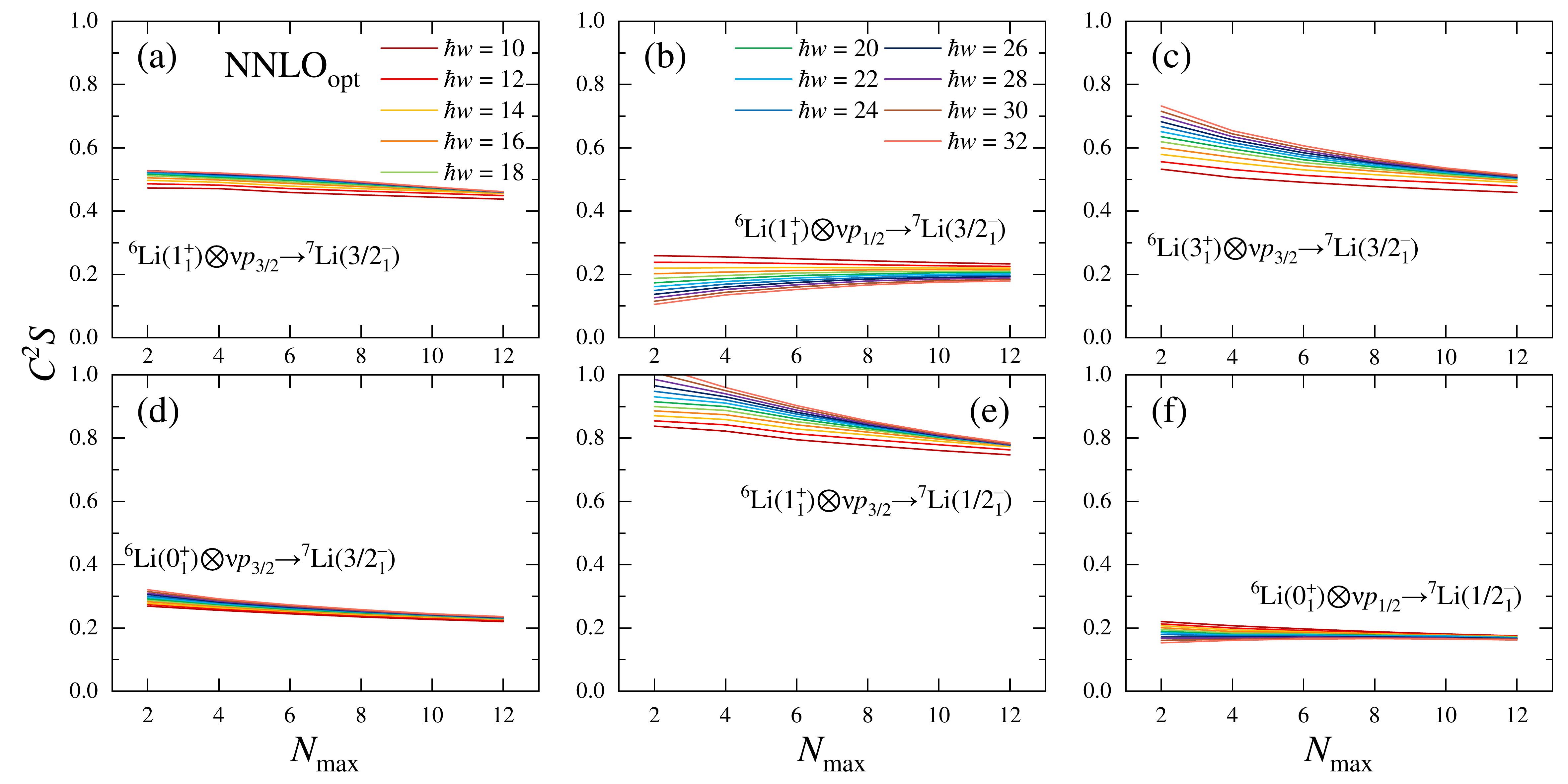}
    \caption{Similar to Fig.~\ref{6He7Li_NNLO}, but for $^6$Li $\otimes$ $p_{3/2,1/2}$ $\to$ $^7$Li using chiral NNLO$_{\rm opt}$ interaction.}
    \label{6Li7Li_NNLO}
\end{figure*}

\begin{figure*}
    \centering
\includegraphics[width=0.75\paperwidth]{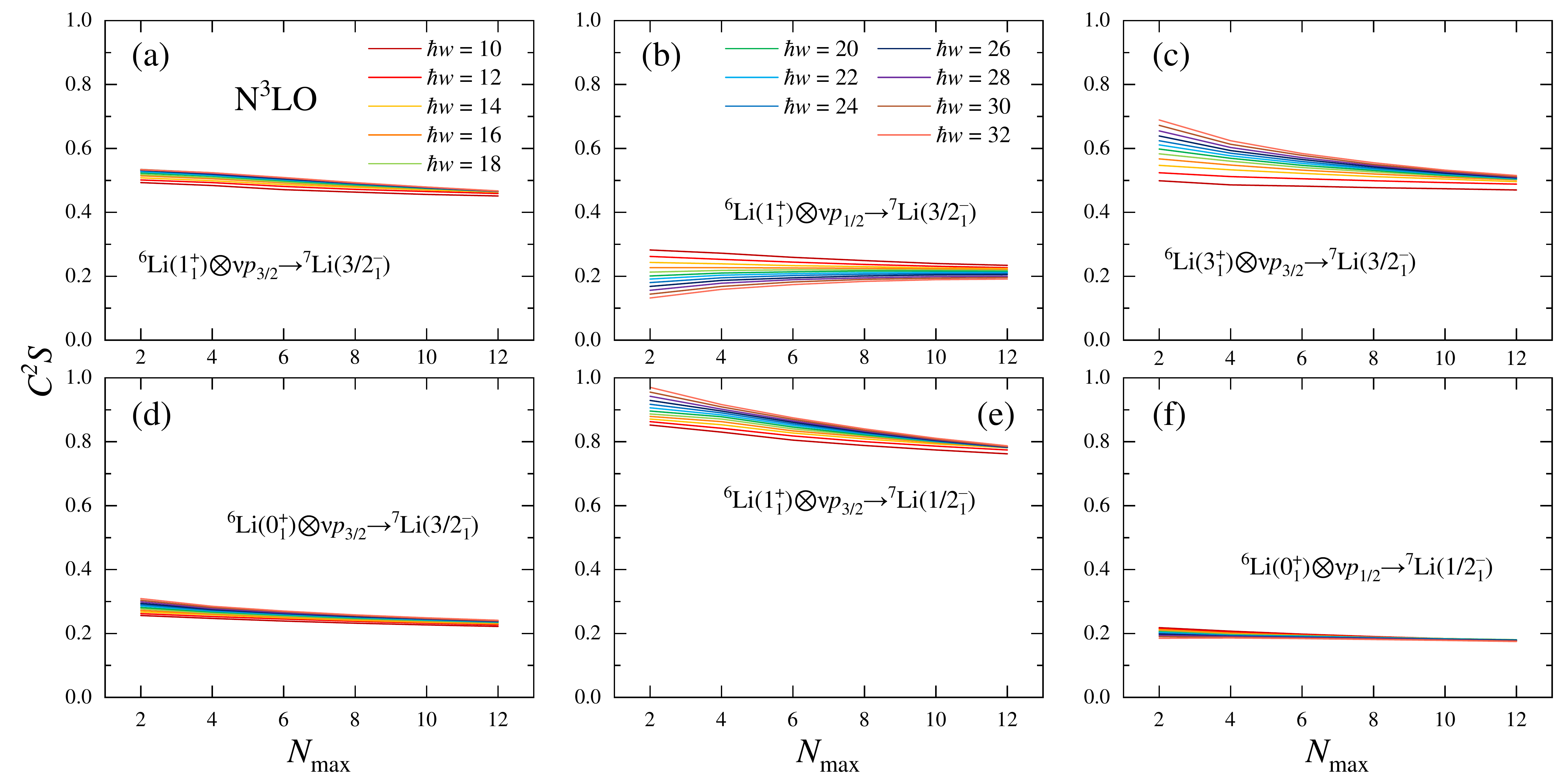}
    \caption{Similar to Fig.~\ref{6He7Li_NNLO}, but for $^6$Li $\otimes$ $p_{3/2,1/2}$ $\to$ $^7$Li using chiral N$^3$LO interaction renormalized via  SRG with $\lambda$ = 2.0 fm$^{-1}$.}
    \label{6Li7Li_SRG}
\end{figure*}

Our NCSM calculation for the energy aligns consistently with preceding NCSM studies~\cite{BARRETT2013131, PhysRevC.86.031301, PhysRevC.79.014308, PhysRevC.69.014311}.
Of particular intrigue is the convergence trend of SF within NCSM calculations, as comprehensive studies specifically addressing SF convergence are notably scarce.
The SF presents a challenge to \textit{ab initio} calculations. The complexity arises because SFs are profoundly sensitive to the intricacies of the nuclear Hamiltonian and the nuances of the many-body wave function~\cite{PhysRevLett.103.202502, PhysRevLett.107.032501}.
Following similar calculations in assessing energy convergence, we performed systematic NCSM calculations of the SFs in $^7$Li using chiral N$^3$LO interaction with $\lambda = 2.0$ $\rm fm^{-1}$ and NNLO$_{\rm opt}$ interactions. The results are depicted in Figs.~\ref{6He7Li_NNLO}, \ref{6He7Li_SRG}, \ref{6Li7Li_NNLO}, \ref{6Li7Li_SRG} as functions of both the model space truncation $N_{\rm max}$ and the frequency $\hbar w$ of HO basis.
Such a presentation is intended to elucidate the convergence behavior of SF in a more detailed manner.
Within the scope of these SF calculations, our primary attention is centered on the $p_{3/2}$ and $p_{1/2}$ partial waves, given their dominant roles in the $A=7$ isotopes.

Let us examine the SF for $^6$He $\otimes$ $p_{3/2,1/2}$ $\to$ $^7$Li using the NNLO$_{\rm opt}$ interaction, as shown in Fig.~\ref{6He7Li_NNLO}. Except for Fig.~\ref{6He7Li_NNLO}(f), the range of the vertical axis of the SFs is $[0:1]$. It is observed that the SF of $^6$He$(0_1^+)$ $\otimes$ $p_{3/2}$ $\to$ $^7$Li$(3/2_1^-)$ is larger than that of other channels. In addition, results reveal a pronounced dependence of the calculated SFs on the HO frequency $\hbar w$ in small model spaces. However, as $N_{\rm max}$ increases, this sensitivity to $\hbar w$ weakens.
As seen in Fig.~\ref{6He7Li_NNLO} (a), the SF for $^6$He$(0_1^+)$ $\otimes$ $p_{3/2}$ $\to$ $^7$Li$(3/2_1^-)$ is almost constant when $N_{\rm max}$ increases, making it difficult to ascertain its convergence results. 
The trend is shown in Fig.~\ref{6He7Li_NNLO} (f). The N$^3$LO result also exhibits a similar trend, see Fig.~\ref{6He7Li_SRG} (a) and (f) for detail. Additionally, to further investigate the impact of renormalization on our calculations, we also perform the calculations using N$^3$LO with $\lambda =2.4$ $\rm fm^{-1}$. The results indicate that while the calculated ground state energies exhibit dependence on the SRG parameter, the SFs are almost independent of the SRG parameter. Moreover, the calculated SF is almost identical in the calculation using N$^3$LO and NNLO$_{\rm opt}$ interaction, despite the varying results on the calculated ground state energy. 
A comparable trend is evident in the NCSM calculations for the SFs of $^6$He$(2_1^+)$ $\otimes$ $p_{3/2}$ $\to$ $^7$Li$(1/2_1^-)$.
Upon reviewing the figures, it is evident that the convergence trend for SF calculations with increasing model space differs at a small level for both interactions, and a convergence pattern is not readily apparent. Moreover, these observed trends diverge from established extrapolation formulas for energy~\cite{PhysRevC.86.054002, PhysRevC.79.014308}. Therefore, we cannot straightforwardly adapt the existing formulas for energy extrapolation to calculate SFs within the NCSM calculations.
However, the convergence is quite different for SFs of different states or partial waves.
For the SFs of $^6$He$(2_1^+)$ $\otimes$ $p_{3/2}$ $\to$ $^7$Li$(3/2_1^-)$, $^6$He$(2_1^+)$ $\otimes$ $p_{1/2}$ $\to$ $^7$Li$(3/2_1^-)$, and $^6$He$(0_1^+)$ $\otimes$ $p_{1/2}$ $\to$ $^7$Li$(1/2_1^-)$, 
a converged trend is evident. This trend shows values almost independent of $\hbar w$ and $N_{\rm max}$ in NCSM calculations, especially within large model spaces.
In the NCSM calculation for the SFs of $^6$Li $\otimes$ $p_{3/2,1/2}$ $\to$ $^7$Li, similar situations are obtained. 
The SFs of $^6$Li$(1_1^+,0_1^+)$ $\otimes$ $p_{3/2}$ $\to$ $^7$Li$(3/2_1^-)$ demonstrate a linear trend as $N_{\rm max}$ increases and the convergence SF cannot be reliably guarantee. However, the calculations of SFs of $^6$Li$(1_1^+)$ $\otimes$ $p_{1/2}$ $\to$ $^7$Li$(3/2_1^-)$ and $^6$Li$(0_1^+)$ $\otimes$ $p_{1/2}$ $\to$ $^7$Li$(1/2_1^-)$ exhibit converged trend, which is similar to the calculations in $^6$He$(2_1^+)$ $\otimes$ $p_{3/2}$ $\to$ $^7$Li$(3/2_1^-)$, see Fig.~\ref{6Li7Li_NNLO} and Fig.~\ref{6Li7Li_SRG} for detail.

Generally, the convergence rate of SFs within NCSM calculations using the HO basis is slow. At present, there are no established extrapolation methods available for reliably predicting these values.
Similar to the NCSM calculations of root-mean-square radius~\cite{PhysRevC.95.014306, PhysRevC.79.021303, Shin_2017, PhysRevC.69.014311, PhysRevC.99.054308}, electric quadrupole moment~\cite{PhysRevC.95.014306, PhysRevC.79.021303, PhysRevLett.99.042501, Shin_2017, PhysRevC.69.014311}, and $E2$ transition~\cite{PhysRevC.95.014306, PhysRevLett.99.042501, Shin_2017}, which correspond to the long-range operators, the convergence of SFs also exhibits slow convergence, which is due to the wave function of the last nucleon need to be expanded with more nodes of localized HO single-particle states.

\begin{center}
\begin{table*}[]
\renewcommand{\arraystretch}{1.0}
\caption{The comparison of theoretical and experimental inclusive cross-sections for knockout reactions from the $^7$Li projectile. The NCSM calculations with NNLO$_{\rm opt}$ and N$^3$LO are performed with $\hbar w = 22$ and 18 MeV, respectively.
The $C^2S$ of the SM includes a center-of-mass motion correction of 1.17.}
\label{c2s_1}
\resizebox{1.00\textwidth}{!}{
\begin{tabular}{ccccccccccccccc}
\hline \hline
\multicolumn{2}{c}{}  & \multicolumn{5}{c}{$C^2S$} & $\sigma_{sp}^{\rm VMC}$~\cite{PhysRevC.86.024315} & $\sigma^{\rm NNLO_{opt}}_{\rm th}$ & $\sigma_{\rm th}^{\rm N^3LO}$ & $\sigma_{\rm th}^{\rm SM}$ & $\sigma_{\rm th}^{\rm VMC}$~\cite{PhysRevC.86.024315} &$\sigma_{\rm th}^{\rm GFMC}$ &$\sigma_{\rm exp}$~\cite{PhysRevC.86.024315}\\ \cline{3-7}    
Reaction & Final state &  NCSM$_{\rm{NNLO_{opt}}}$ & NCSM$_{\rm{N^3LO}}$ & SM~\cite{PhysRevC.86.024315} & VMC~\cite{PhysRevC.86.024315} & GFMC~\cite{PhysRevC.84.024319} & (mb) & (mb) & (mb) & (mb) & (mb) & (mb) & (mb)\\ \hline \hline
    ($^7$Li,$^6$Li) &$1^+$ & 0.661 & 0.683 & 0.733 & 0.715 & 0.668 & 56.9(13) & 37.61 & 38.86 & 41.71 & 40.7(9) & 38.00 \\ 
    ($^7$Li,$^6$Li) & $0^+$ & 0.231 & 0.234 & 0.389 & 0.219 & 0.203 & 56.8(26) & 13.12 & 13.29 & 22.10 & 11.9(5) & 11.53\\
     Inclusive & & & & & & & &  50.73 & 52.15 & 63.81 & 52.6(10) & 49.53 & 30.7(18) \\
     ($^7$Li,$^6$He) & $0^+$ & 0.459 & 0.466 & 0.806 & 0.439 & 0.406 &  60.8(31) & 27.91 & 28.33 & 49.00 & 26.7(14) & 24.68 \\ 
     Inclusive & & & & & & & & 27.91 & 28.33 & 49.00 & 26.7(14) & 24.68 & 13.4(7)\\
\hline \hline
\end{tabular}}
\end{table*}
\end{center}

Our NCSM calculations highlight the complexity and unpredictability inherent in the behavior of SF, emphasizing the need for more refined approaches to describe these behaviors effectively. It is worth noting that, although the convergence of SF calculations within the NCSM framework remains uncertain, the method inherently incorporates a significant degree of configuration mixing.
To test the SFs in NCSM calculations, we adapt them to calculate the knockout reaction cross-section.
Although the convergence of SFs calculations in the NCSM is not obtained, the calculated SFs within $N_{\rm max} =12$ model space are adopted in the following knockout reaction calculations. 

The knockout reaction ($^7$Li,$^6$He) and ($^7$Li,$^6$Li) have been comprehensively analyzed in Ref.~\cite{PhysRevC.86.024315}, where SFs derived from both the SM and \textit{ab initio} VMC methods were utilized to calculate the cross-sections for neutron and proton knockout from a $^7$Li beam, which offer valuable theoretical and experimental benchmarks. The calculated SFs from NCSM are also employed in the above knockout reaction calculations.
It is important to note that the $\rm NCSM_{NNLO_{opt}}$ analysis uses a calculation with $\hbar w = 22$ MeV, while the $\rm NCSM_{N^3LO}$ employs results from $\hbar w = 18$ MeV.
Additionally, for comparison purposes, we also consider SFs derived from the \textit{ab initio} Green’s Function Monte Carlo (GFMC) method~\cite{PhysRevC.84.024319}.
The $\sigma_{sp}^{\rm VMC}$, SFs of both SM and VMC, and $\sigma_{\rm exp}$ presented in Table~\ref{c2s_1} are taken from Ref.~\cite{PhysRevC.86.024315}. 
To evaluate the effects of calculated SF on cross-section calculations, the same $\sigma_{sp}$ derived from the VMC wave function is adopted~\cite{PhysRevC.86.024315}. The $\sigma_{sp}$ are combined with SFs from the SM, NCSM, VMC, and GFMC in the calculations of cross-section, as depicted in Eq.~(\ref{eq-sigma}).



Interestingly, in the NCSM approach, the two interactions yield nearly identical results, revealing minimal deviations, which is due to the calculated SF from NCSM using that two interactions are almost identical.
Moreover, the SFs from our NCSM calculations align closely with those from VMC and GFMC.
However, the SFs calculated via SM are larger than those of \textit{ab initio} NCSM, VMC, and GFMC calculations.
This disparity is especially evident in the SFs of $^6$Li$(0_1^+)$ $\otimes$ $p_{3/2}$ $\to$ $^7$Li$(3/2_1^-)$ and $^6$He$(0_1^+)$ $\otimes$ $p_{3/2}$ $\to$ $^7$Li$(3/2_1^-)$.

For $^7$Li single proton knockout reaction, SFs derived from the NCSM give an inclusive cross-section of approximately 28 $\rm mb$, which aligns well with the \textit{ab initio} VMC and GFMC calculations, although these results tend to overestimate compared to experimental data. In contrast, the SM gives a notably higher cross-section of 49 $\rm mb$. A similar pattern is observed in the $^7$Li single neutron knockout reaction,  where the SM calculations also overestimate the cross-section compared to experimental data, and the calculations are improved within the \textit{ab initio} frameworks of NCSM, VMC, and GFMC calculations.


The discrepancy between theoretically predicted and experimentally measured reaction cross-sections has long been a subject of intensive scrutiny in nuclear physics. 
Despite numerous studies and investigations into this inconsistency, a comprehensive explanation for these deviations remains elusive.
Such discrepancies can often hint at overlooked or simplified physics in theoretical frameworks.
Undoubtedly, the elucidation of these discrepancies requires a combined effort, leveraging insights from both reaction and structural models.
In reaction modeling, the widespread adoption of adiabatic (or sudden) and eikonal approximations~\cite{PhysRevLett.130.172501,doi:10.1146/annurev.nucl.53.041002.110406} may introduce potential inaccuracies. 
Delving deeper into nuclear structure reveals that even models providing structural inputs to reaction frameworks have inherent limitations. 
The intrinsic complexity of the NCSM provides it a unique advantage: cross-sections derived from this model show a better alignment with experimental observations compared to that of SM calculations in some cases.
Furthermore, NCSM calculations align well with results from other state-of-the-art theoretical approaches, such as the \textit{ab initio} VMC and GFMC. This confluence of results from various models shows that NCSM can be used in practice to evaluate cross-sections, but only from an empirical point of view. A sound determination of cross-sections in NCSM indeed demands SFs to be independent of the used model space and, hence to be converged or renormalized in a precise manner, which is not yet the case.

\section{summary}

We undertook comprehensive calculations of spectroscopic factors using the \textit{ab initio} NCSM for $^7$Li, utilizing two distinct chiral interactions: NNLO$_{\rm opt}$ and N$^3$LO.
In the NCSM, calculated energies demonstrate good convergence with both the increasing model space and the varying HO frequency $\hbar w$. 
Contrary to energy calculations, SFs display a relatively slow convergence as the model space expands. This slow convergence complicates the application of existing energy convergence formulas, making the predictability and reliability of SF calculations more challenging. This absence of a pattern necessitates the development of more nuanced approaches or methodologies to assess and ensure the convergence of SFs in theoretical models.
The calculated SFs are also employed in the knockout reaction cross-section calculations.
It is noteworthy that the results of NCSM calculations are consistent with the results from both \textit{ab initio} VMC and GFMC methods. Nevertheless, this agreement is still not fully justified from a theoretical point of view due to the absence of convergence of SFs in NCSM.

\textit{Acknowledgments.}~
 This work has been supported by the National Key R\&D Program of China under Grant No. 2023YFA1606403; the National Natural Science Foundation of China under Grant Nos.  12205340, 12175281, and 12121005;  the Gansu Natural Science Foundation under Grant No. 22JR5RA123;  the Strategic Priority Research Program of Chinese Academy of Sciences under Grant No. XDB34000000; the Key Research Program of the Chinese Academy of Sciences under Grant No. XDPB15; the State Key Laboratory of Nuclear Physics and Technology, Peking University under Grant No. NPT2020KFY13. The numerical calculations in this paper have been done on  Hefei advanced computing center.

\bibliography{Ref}

\end{document}